\documentclass[amsmath,amssymb,superscriptaddress,aps,showpacs,10pt,onecolumn]{revtex4-1}

\usepackage{amsfonts}
\usepackage{graphicx}
\usepackage{soul}

\newcommand{\kb}[1]{\textrm{\tiny{#1}}}

\newcommand{\kd}[1]{\mathbf{#1}}

\begin{document}

\title{Ultralong-range Rydberg molecules in combined electric and magnetic fields}

\date{\today}
%

\author{Markus Kurz}
\affiliation{Zentrum f\"ur Optische Quantentechnologien, Luruper Chaussee 149, 22761 Hamburg, Universit\"at Hamburg, Germany}
\author{Peter Schmelcher}
\affiliation{Zentrum f\"ur Optische Quantentechnologien, Luruper Chaussee 149, 22761 Hamburg, Universit\"at Hamburg, Germany}
\affiliation{The Hamburg Centre for Ultrafast Imaging, Luruper Chaussee 149, 22761 Hamburg, Universit\"at Hamburg, Germany}

\begin{abstract}
We investigate the impact of combined electric and magnetic fields on the structure of ultralong-range polar Rydberg molecules. Our focus is hereby on the parallel as well as the crossed field configuration taking into account both the $s$-wave and $p$-wave interactions of the Rydberg electron and the neutral ground state atom. We show the strong impact of the $p$-wave interaction on the ultralong-range molecular states for a pure $B$-field configuration. In the presence of external fields the angular degrees of freedom acquires vibrational character and we encounter two- and three-dimensional oscillatory adiabatic potential energy surfaces for the parallel and crossed field configuration, respectively. The equilibrium configurations of local potential wells can be controlled via the external field parameters for both field configurations depending of the specific degree of electronic excitation. This allows to tune the molecular alignment and orientation. The resulting electric dipole moment is in the order of several kDebye and the rovibrational level spacings are in the range of $2-250$ MHz. Both properties are analyzed with of varying field strengths. 
\end{abstract}
%
%
\maketitle
\section{Introduction \label{int}}
The preparation and control of ultracold atomic and molecular systems offer a unique platform for a detailed understanding and analysis of fundamental quantum processes. The external degrees of freedom can be accessed by designing and switching almost arbitrarily shaped traps \cite{Grimm00,Folman02,Pethick08} which are created by external electric, magnetic or electromagnetic fields. Furthermore, interatomic interaction strength can be tuned via magnetic or optical Feshbach resonances \cite{Koehler06,Bloch08,Chin10,Jochim03,Knoop09,Repp13}.
A prominent new species are the weakly bound ultralong-range diatomic molecules composed of a ground state and a Rydberg atom. Theoretically predicted more than a decade ago \cite{Greene00} they have been recently discovered experimentally \cite{Bendkowsky09}. The molecular Born-Oppenheimer potential energy surfaces, which are responsible for the diatomic binding, show for these species an unusual strong oscillatory behavior providing a large number of local potential minima. This structure can be understood intuitively as the interaction of a neutral ground state atom with the Rydberg atom which can be modeled as a low energy scattering potential between the quasi free Rydberg electron and the ground state atom. In a first approximation, this so called $s$-wave Fermi pseudopotential ansatz \cite{Fermi34,omont77} leads to diatomic molecular species with unique features like internuclear distances in the order of the Rydberg atom. For the considered quantum numbers $n \approx 30$-$50$ the vibrational binding energies are in the MHz and GHz regime depending on the type of states. More specifically, low-angular momentum non-polar states and large angular momentum polar states, so-called trilobites, have been predicted \cite{Greene00}. They possess electric dipole moments in the range of 1 Debye (low-$\ell$) up to 1 kDebye (high-$\ell$) in the polar case. The large electric dipole moment of the latter make them accessible for external field  manipulation which opens the possibility for the control of the molecular degrees of freedom. For instance, the selective excitation of stationary molecular $D$-states in external magnetic fields was demonstrated recently possessing an extraordinary degree of alignment or antialignment with respect to the magnetic field axis \cite{Krupp14}.
The impact of either magnetic or electric fields on ultralong-range molecules has been studied in previous works \cite{Lesanovsky07,Kurz13}. It has been shown that in the presence of an external field the angular degree of freedom between the field and the internuclear axis acquires vibrational character resulting in two-dimensional adiabatic potential energy surfaces (PES). A pure magnetic field yields in molecular states oriented perpendicular to the molecular axis and leads, with increasing field strength, to a monotonic lowering of the magnitude of the electric dipole moment \cite{Lesanovsky07}. However, a pure electric field forces the molecule into a parallel oriented configuration with an electric dipole of growing magnitude for increasing electric field strength \cite{Kurz13}. Combining electric and magnetic fields in a crossed field configuration the existence and properties of so-called giant dipole ultralong-range molecules had been analyzed \cite{Kurz12} which corresponds to the situation of a magnetic field dominating over the Coulomb interaction.

In the present work we explore the impact of combined electric and magnetic fields, more specifically the cases of parallel and crossed (perpendicular) fields, on the structure and dynamics of high-$\ell$ ultralong-range diatomic rubidium molecules. Our analysis goes beyond the $s$-wave approximation taking into account the next order $p$-wave term in the Fermi-pseudopotential. Already for the case of a magnetic field only we demonstrate the strong impact of the $p$-wave contribution. Due to $p$-wave interactions the potential wells providing the weakly bound trilobite states vanish beyond a critical magnetic field strength and consequently no bound states exist anymore. For combined electric and magnetic fields the topology of the PES strongly depend on the specific field configuration and the applied field strengths. The resulting PES show a strong oscillatory behavior with depths up to hundreds of MHz and we find rovibrational bound states with level spacings in the MHz regime. By tuning the field parameters separately we can control the molecular orientation for the parallel configuration 
from a perpendicular to an antiparallel molecular configuration with respect to the magnetic field.
For the crossed field configuration the alignment can be tuned from an aligned to an antialigned molecular state. For both field configurations we present an analysis of the electric dipole moment.

In detail we proceed as follows. In Section \ref{mol_ham_int} we present the molecular Hamiltonian and a discussion of the underlying interactions. Section \ref{method} and \ref{p_wave} contain the methodology and our results of the pure magnetic field configuration, respectively. 
In Section \ref{pot_ener_surf} we analyze the impact of the combined fields on the topology of the PES for the parallel as well as the crossed field configuration. Their rovibrational spectra are addressed in Section \ref{rovi_states}. A detailed study of the alignment and orientation as well as the corresponding electric dipole moment are provided in Sections \ref{mol_align_orien} and \ref{elec_mag}, respectively. Finally, Section \ref{conclusion}  contains our conclusions.    
\section{Molecular Hamiltonian and Interactions \label{mol_ham_int}}
We consider a highly excited Rydberg atom interacting with a ground state neutral atom, often called the 'perturber' atom in combined static and homogeneous electric and magnetic fields. Our focus will be on the $^{87}$Rb atom. The Hamiltonian treating the Rb ionic core and the neutral perturber as point particles is given by (if not stated otherwise, atomic units will be used throughout)
\begin{eqnarray}
H=\frac{\kd{P}^{2}}{M}+H_{\rm{el}} +V_{\textrm{n,e}}(\kd{r},\kd{R}),\label{ham}\ \
H_{\rm{el}}=H_0 + \kd{E}\kd{r} + \frac{1}{2}\kd{B}\kd{L}+\frac{1}{8}(\kd{B}\times\kd{r})^2,\ \ H_0=\frac{\kd{p}^2}{2m_{\kb{e}}} + V_{l}(r), 
\end{eqnarray}
where $(M,\kd{P},\kd{R})$ denote the atomic $Rb$ mass and the relative momentum and position of the neutral perturber with respect
to the ionic core. $(m_{\rm{e}},\kd{p},\kd{r})$ indicate the corresponding quantities for the Rydberg electron. The electronic Hamiltonian $H_{\rm{el}}$ consists of the field-free Hamiltonian $H_0$ and the usual Stark, Zeeman and diamagnetic terms of an 
electron in static external $\kd{E}$-/$\kd{B}$-fields. 
$V_l(\kd{r})$ is the angular momentum-dependent one-body pseudopotential felt by the valence electron when interacting 
with the ionic core \cite{Marinescu94}. For low-lying angular momentum states the electron penetrates the finite ionic $\text{Rb}^{+}$-core 
which leads to a $\ell$-dependence of the interaction potential $V_l(r)$ due to polarization and scattering effects 
\cite{Gallagher}. Throughout this work we choose the direction of the magnetic field to coincide with the z-axis of the 
coordinate system, i.e.\ $\kd{B}=B\kd{e}_{z}$. Finally, the interatomic potential $V_{\rm{n},e}$ for the 
low-energy scattering between the Rydberg electron and the neutral perturber is described as a so-called Fermi-pseudopotential 
\begin{eqnarray}
V_{\textrm{n,e}}(\kd{r},\kd{R})&=&2\pi A_{s}[k(R)]\delta(\kd{r}-\kd{R})+6\pi  A^{3}_{p}[k(R)]\overleftarrow{\nabla} \delta(\kd{r}-\kd{R}) \overrightarrow{\nabla}.\label{inter}
\end{eqnarray}
Here we consider the triplet ($S=1$) scattering of the electron from the spin-$\frac{1}{2}$ ground state alkali atom. 
Suppression of singlet scattering events can be achieved by an appropriate preparation of the initial atomic gas. 
In Eq.\ (\ref{inter}) $A_{s}[k(R)]=-\tan(\delta_0(k))/k$ and $A^{3}_{p}[k]=-\tan(\delta_1(k))/k^{3}$ denote the energy-dependent 
triplet $s$- and $p$-wave scattering lengths, respectively, which are evaluated from the corresponding phase shifts 
$\delta_l(k),\ l=0,1$. The kinetic energy $E_{\rm{kin}}=k^2/2$ of the Rydberg electron at the collision point with the 
neutral perturber can be taken according to $k^2 / 2=1/R-1/2n^2$, which represents a semiclassical approximation.
\begin{figure}[hbt]
\includegraphics[width=0.3\linewidth]{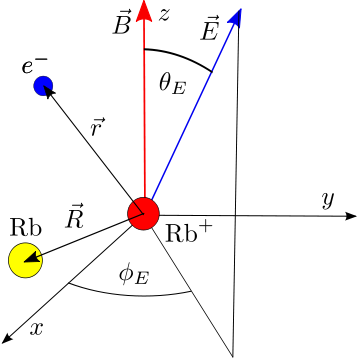}
\caption{(Color online) A sketch of the considered setup. An ultralong-range Rydberg molecule is exposed to external electric $\kd{E}$ and magnetic $\kd{B}$-fields. The molecule consists of a rubidium Rydberg atom (Rb$^{+}$ ionic core plus valence electron ($e^{-}$)) and a neutral ground state atom (Rb) which interact via a low energy electron-atom scattering potential $V_{\textrm{n,e}}(\kd{r},\kd{R})$). The $\kd{B}$-field points along the $z$-axis while the direction of the $E$-field is specified by the angles ($\theta_{\rm{E}},\phi_{\rm{E}}$).} 
\label{setup} 
\end{figure}
\section{Methodology \label{method}}
In order to solve the eigenvalue problem associated with the Hamiltonian \eqref{ham} we adopt an adiabatic ansatz for 
the electronic and heavy particle dynamics. We write the total wave function as $\Psi(\kd{r},\kd{R})=\psi(\kd{r}; \kd{R} )\phi(\kd{R})$ and 
obtain within the adiabatic approximation
\begin{eqnarray}
[H_0 + \kd{E}\kd{r}+ \frac{1}{2}\kd{B}\kd{L}+\frac{1}{8}(\kd{B}\times\kd{r})^2 +V_{\kb{n,e}}(\kd{r},\kd{R})]\psi_{i}(\kd{r};\kd{R})&=&\epsilon_{i}(\kd{R})\psi_{i}(\kd{r};\kd{R}), \label{hamelec}\\
(\frac{\kd{P}^{2}}{M}+\epsilon_{i}(\kd{R}))\phi_{ik}(\kd{R})&=&\mathcal{E}_{ik}\phi_{ik}(\kd{R}),\label{hamrovi}
\end{eqnarray}
where $\psi_i$ describes the electronic molecular wave function for a given 
relative position $\kd{R}$ and $\phi_{ik}$ determines the rovibrational state of the molecule. To calculate the potential 
energy surface $\epsilon_{i}(\kd{R})$ we expand $\psi(\kd{r};\kd{R})$ in the eigenbasis of $H_0$, i.e.\ $\psi_i(\kd{r};\kd{R})=\sum_{nlm} C^{(i)}_{nlm}(\kd{R})\chi_{nlm}(\kd{r})$ with 
$H_0\chi_{nlm}(\kd{r})=\varepsilon_{nl}\chi_{nlm}(\kd{r}),\ \chi_{nlm}(\kd{r})\equiv \langle\kd{r}|nlm \rangle=R_{nl}(r)Y_{lm}(\vartheta,\varphi)$. 
For $l \ge l_{\rm{min}}=3$ we neglect all quantum defects, i.e.\ $H_0$ is identical to the hydrogen problem. 
Finally, we have to solve the following eigenvalue problem
\begin{eqnarray}
(\varepsilon_{nl}-\epsilon(\kd{R})+m\frac{B}{2})C_{nlm}+\sum_{n'l'm'}C_{n'l'm'}(\langle nlm |Er\cos(\Omega)+\frac{B^2}{8}r^2\sin^2(\vartheta)|n'l'm \rangle
+\langle nlm |V_{\textrm{n,e}}(\kd{r},\kd{R})|n'l'm' \rangle) = 0,
\end{eqnarray}
with $\cos(\Omega)=\sin(\theta_E)\sin(\vartheta)\cos(\phi_E-\varphi)+\cos(\theta_E)\cos(\vartheta)$. The angles ($\theta_E,\phi_E$) specify the direction of the electric field (see Fig.\ \ref{setup}). In this work we analyze the parallel ($\theta_E=\phi_E=0$) and perpendicular ($\theta_E=\pi/2,\ \phi_E=0$) field configuration. To study the different configurations we use standard numerical techniques for the diagonalization of the resulting hermitian matrices. Throughout this work 
we focus on the high-$\ell$ $n=35$ manifold, for other $n$-quantum numbers the underlying physical processes remain similar. In case of zero electric field these high-$\ell$ manifold provides the trilobite states \cite{Greene00}. To ensure convergence we vary the number of basis states finally achieving a relative accuracy 
of $10^{-3}$ for the energy. For the $n=35$ trilobite manifold we used, in addition to the degenerate 
$n=35,\ l \ge 3$ manifold, a basis set that includes the $38s,\ 37d,\ 36p$ quantum defect split states due to their
energetical closeness. This basis set contains $1225$ states in total.

From Eqs.\ (\ref{hamelec}) and (\ref{hamrovi}) we already deduce some symmetry properties of the states $\psi,\ \phi$ 
and the energies $\epsilon$ for the different field configurations. If $P_{\kd{r},\kd{R},\kd{E}}$ denotes the generalized parity operator that transforms 
$(\kd{r},\kd{R},\kd{E})\rightarrow(-\kd{r},-\kd{R},-\kd{E})$ we have 
$[H,P_{\kd{r},\kd{R},\kd{E}}]=[V_{\rm{n,e}}(\kd{r},\kd{R}),P_{\kd{r},\kd{R},\kd{E}}]=0$. This means that the states $\Psi,\ \psi$ and $\phi$  are parity (anti)symmetric and the PES fulfill $\epsilon_{||,\perp}(\kd{R};\kd{E})=\epsilon_{||,\perp}(-\kd{R};-\kd{E})$ where ($||,\perp$) denote the PES in case of parallel and perpendicular fields, respectively. In addition, if $\theta_E=0$ (parallel configuration), the PES possess an azimuthal symmetry, e.g.\ $\epsilon_{||}(\kd{R})=\epsilon_{||}(R,\theta)$ and the vector defining the internuclear axis can, without loss of generality, be chosen to lie in the $x$-$z$-plane. In contrast, if $\theta_E=\pi/2$ (perpendicular configuration), the PES depend on the azimuthal coordinate $\phi$ as well and possess only reflection symmetries with respect to the $x$-$y$-plane and the $x$-$z$-plane, i.e.\ $\epsilon_{\perp}(R,\pi-\theta,\phi)=\epsilon_{\perp}(R,\theta,\phi)$ and $\epsilon_{\perp}(R,\theta,2\pi-\phi)=\epsilon_{\perp}(R,\theta,\phi)$. In this work the energy offset of all PES is the dissociation limit of the atomic states Rb($5s$)+Rb($n=35,\ l \ge 3$).  
\section{$p$-wave interaction effects in a magnetic field \label{p_wave}}
Before we study the combined field configurations let us analyze the system for zero electric field ($E=0$). In \cite{Lesanovsky07} has been done considering only the $s$-wave scattering for the electron-perturber interaction. In contrast to this we here include also the $p$-wave interaction.

\begin{figure}[hbt]
\includegraphics[width=0.5\linewidth]{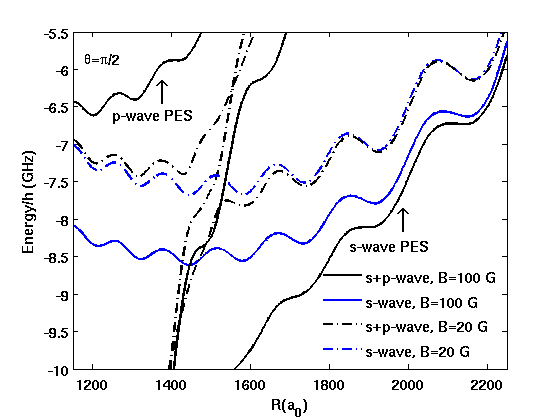}
\caption{(Color online) Comparison between the pure $s$-wave scattering dominated potential curves (blue lines) and the combined $s$- and $p$-wave potential curves (black lines). Provided are cuts for $\theta=\pi/2$ for different magnetic field strengths $B=20$ G (dashed lines) and $B=100$ G (solid lines).}
\label{avoided} 
\end{figure}

Fig.\ \ref{avoided} shows an one-dimensional cut through the $s$- and $p$-wave dominated PES (black lines) in comparison with the purely $s$-wave dominated potential curves (blue lines) for a magnetic field strength $B=20$ G (dashed lines) and $100$ G (solid lines). The specific cut is taken along the $\theta=\pi/2$ direction. As discussed in \cite{Lesanovsky07} in the case of a pure $s$-wave scattering potential the field-dependent terms represent a perturbation with respect to the field-free molecular Hamiltonian and the considered potential curve is just the known trilobite potential curve \cite{Greene00} shifted by the Zeeman splitting. The resulting PES provide, beside a global minimum between $R=1400a_0$ and $1500a_0$, a number of local minima which are taken on for the  $\theta=\pi/2$ configuration for which the internuclear axis is perpendicular to the applied field. This behavior is clearly visible in Fig.\ (\ref{avoided}) for the blue curves which represent the purely $s$-wave dominated potential curves.
However, including the $p$-wave scattering term changes the situation substantially. As discussed in detail in \cite{Kurz13} in the field-free case additional potential curves arise causing avoided crossing in the vicinity of the global minimum of the $s$-wave trilobite curve. As shown in Fig.\ \ref{avoided} for $\theta=\pi/2$ we are faced with two additional potential curves. For $B=20$ G (dashed, black curves) the avoided crossings in the region $R=1400a_0-1600a_0$ known from the field-free case are still visible. With decreasing energetically order we first have an oscillating potential curve with a potential minimum at approximately $-7.44$ GHz for $R=1312a_0$ that strongly increases for $R \ge 1400a_0$. Because it arises from the additional $p$-wave interaction we denote this PES as {\it p-wave PES}. Second, we find a monotonically increasing potential curve ranging from $R=1400a_0$ to $1500a_0$. In this work this PES is not of interest because it does not exhibit any potential minima and therefore contains no bound states. The third and energetically lowest potential curve is the one providing the ultralong-range molecules ("trilobite states") from \cite{Greene00} in the field-free case. This curve does not possess a global minimum any more. It monotonically increases till $R \approx 1450a_0$ and thereafter possesses an oscillatory behavior with local potential wells of depths in the hundred MHz regime. We observe that for increasing radial distance $R$ the $s$-wave character becomes more and more dominant. This curve provides metastable bound states. Although this potential curve is at least in a certain region already strongly $p$-wave interaction affected in the field-free case \cite{Kurz13} we denote it as the {\it s-wave PES}.

Finally, the effect of an increasing field strength on the $p$-wave dominated potential curves can be seen in Fig.\ \ref{avoided} as well. We present the $p$-wave dominated potential energy curves for a magnetic field strength of $B=100$ G (solid black curves). An obvious consequence is that the $s$- and $p$-wave PES have moved up and down in energy, respectively, while the energetically intermediate curve is still monotonically increasing, but now in the enlarged spatial region $1400a_0 \le R \le 1600a_0$. This behavior can be understood by applying perturbation theory. For $0\le B \le 50$ G the PES are well reproduced by the expression   
\begin{eqnarray}
\epsilon^{(s,p)}_{\rm{per}}(\kd{R};B)=\epsilon^{(s,p)}_{0}(\kd{R})+\frac{B}{2}\langle (s,p);\kd{R} | L_{z}| (s,p);\kd{R} \rangle + \frac{B^2}{4}\sum_{n \not= (s,p)}\frac{|\langle n;\kd{R} |L_z | (s,p);\kd{R} \rangle|^2}{\epsilon^{(s,p)}_{0}(\kd{R})-\epsilon^{(n)}_{0}(\kd{R})},\label{pert}
\end{eqnarray} 
where $| n;\kd{R} \rangle$ and $\epsilon^{(n)}_{0}(\kd{R})$ denote the field-free adiabatic electronic eigenstates and eigenenergies. The diamagnetic term in (\ref{ham}) can be neglected here. Obviously, the term of $O(B^2)$ potentially becomes relevant in the region of avoided crossings of the field-free curves which are localized around $R \approx 1450a_0$ \cite{Kurz13}. For increasing magnetic field strength the term proportional to $B^2$ in (\ref{pert}) becomes dominant in spatial regions beyond the point of the field-free avoided crossings ($R \approx 1450a_0$). This causes the PES to separate energetically in the way as it can be seen in Fig. \ref{avoided} for the field strength $B=20$ G and $B=100$ G. Besides the energetic separation a second consequence of the $p$-wave interaction is the disappearance of the local potential wells in case of the $s$-wave curve with increasing magnetic field strength.

Indeed for $B=20$ G and $R \ge 1450a_0$ we find four local potential wells with depths of the order of hundreds of MHz whereas for $B=100$ G no local potential wells are present. Instead the $s$-wave PES monotonically increases and possesses two plateaus at radial positions where the former two outermost potential wells had been localized. Therefore in case of a pure magnetic field and beyond a critical field strength of $B_{\rm{cr}}=100$ G the $s$-wave curve does  
not provide any bound states for the $\theta=\pi/2$ configuration.\\
\begin{figure}[hbt]
\begin{minipage}{0.48\textwidth}
\includegraphics[width=\linewidth]{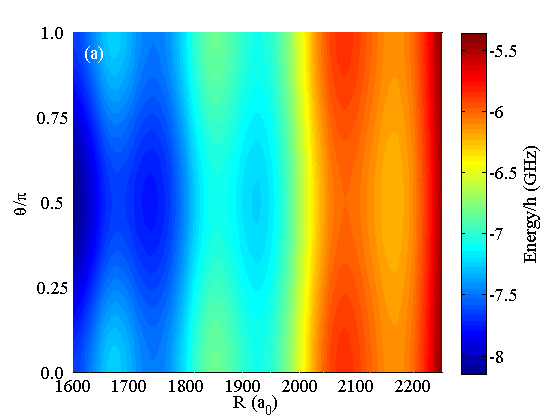}
\end{minipage}
\begin{minipage}{0.48\textwidth}
\includegraphics[width=\linewidth]{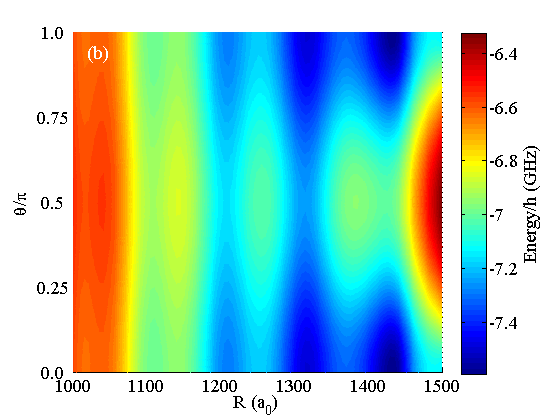}
\end{minipage}
\caption{(Color online) (a) The $s$-wave PES for $B=40$ G for zero electric field showing a reflection symmetry with respect to $\theta=\pi/2$. (b) The $p$-wave PES for $B=40$ G for zero electric field. It possesses a global minimum at $R=1432a_0$ and $\theta=0,\pi$. We clearly see how the region around $\theta=\pi/2$ is affected by the level repulsion with respect to the $s$-wave PES.}
\label{p_wave_B_comp_par}
\end{figure}
In Fig.\ \ref{p_wave_B_comp_par}(a) the complete two-dimensional $s$-wave PES is shown as a function of ($R,\theta$) for $B=40$ G for radial distances $1600a_0 \le R \le 2250a_0$. We observe a $\theta \rightarrow \pi-\theta$ reflection symmetric potential surface with local potential minima at $R_{\rm{eq}}=1728a_0,\ 1918a_0,\ 2159a_0$ and $\theta_{\rm{eq}}=\pi/2$.

In Fig.\ \ref{p_wave_B_comp_par}(b) the complete two-dimensional $p$-wave PES is shown as a function of ($R,\theta$) for $B=40$ G for radial distances $1000a_0 \le R \le 1500a_0$. We observe a potential surface with the global equilibrium positions at $R_{\rm{eq}}=1432a_0,\ \theta_{\rm{eq}}=0,\pi$. This $p$-wave PES provides   bound rovibrational states. The region around $R=1500a_0,\ \theta=\pi/2$ is strongly affected by the level repulsion of the $s$- and $p$-wave PES as it has been described above. In Fig.\ \ref{p_wave_B_comp_par}(a) and \ref{p_wave_B_comp_par}(b) we see also that in this region the $s$- and $p$-wave PES strongly decreases and increases, respectively. 

However, for $R=1500a_0$ and $\theta$ approaching $\pi$ or $0$ respectively, the effect of the $s$- and $p-$wave level repulsion decreases for the $s$- and $p$-wave PES. For $\theta=0,\pi$ this effect vanishes completely, which is due to the fact that the Hamiltonian (\ref{hamelec}) then separates into a $m=0$ and $|m|=1$ block. The considered $s$- and $p$-wave curves arise due to the diagonalization of the $m=0$ subspace of the electronic problem (\ref{hamelec}). If we neglect the diamagnetic term in (\ref{hamelec}) the Zeeman interaction term does not couple the $s$- and $p$-wave curves because $L_{z}| n,l,0 \rangle=0,\ \forall n,l$. Because of this the topology of the PES is unaffected by the applied magnetic field for $\theta=0,\pi$. 
\section{Potential energy surfaces \label{pot_ener_surf}}
\subsection{Parallel field configuration}
Let us begin with the analysis of the s-$wave$ PES. We focus on the regime of field strengths $B=0-100$ G and $E=0-100$ $\frac{V}{m}$. As presented in \cite{Greene00} the potential $s$-wave PES is in the absence of any external field independent of $\theta$. For a finite electric or magnetic field strength, this spherical symmetry is broken, which has been discussed in refs.\ \cite{Lesanovsky07, Kurz13}. In case of a vanishing electric field ($\kd{B}=B\kd{e}_z,\ 0 < B < B_{\rm{cr}},\ E=0$) the $s$-wave PES provides a number of local minima which are taken on for the perpendicular configuration $\theta=\pi/2$. In case of an electric field only ($E=E\kd{e}_z, E > 0,\ B=0$) it forces the electron density to align in its negative direction, which leads to a higher density in the $-z$-direction. For this reason we find the global minimum of the $s$-wave PES for the antiparallel field configuration at $\theta=\pi$.

\begin{figure}[hbt]
\includegraphics[width=0.5\linewidth]{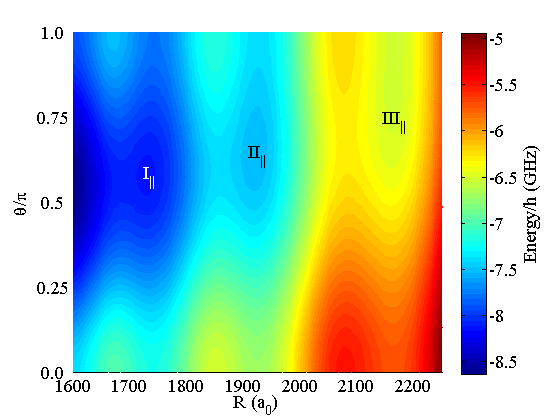}
\caption{(Color online) $s$-wave PES as a function of $(\theta,R)$ ($\theta_{\rm{E}}=\phi_{\rm{E}}=0$, $B=60$ G, $E=20\ \frac{V}{m}$). We observe three local potential wells which are labeled as ($\text{I}_{\rm{||}}$), ($\text{II}_{\rm{||}}$)  and ($\text{III}_{\rm{||}}$). The wells $\text{II}_{\rm{||}}$ and $\text{III}_{\rm{||}}$ provide rovibrational bound states with a level spacing of $2$-$20$ MHz.}
\label{B_60_E_20_exc_10} 
\end{figure}

In case of finite parallel electric and magnetic field strengths we can tune the topology of the $s$-wave PES between the pure electric and magnetic field limits. To be specific we choose $B=60$ G and vary the electric field strength $E=0-100$ $\frac{V}{m}$. Fig.\ \ref{B_60_E_20_exc_10} presents the $s$-wave PES for $B=60$ G, $E=20\ \frac{V}{m}$. We observe three local potential minima which we label with ($\text{I}_{\rm{||}}$), ($\text{I}_{\rm{||}}$)  and ($\text{III}_{\rm{||}}$). In case of $E=0$ we find these potential wells along the perpendicular configuration ($\theta=\pi/2$) with the radial minima position at $R_{\rm{I}_{||}}=1728a_0$, $R_{\rm{II}_{||}}=1918a_0$ and $R_{\rm{III}_{||}}=2159a_0$ respectively (see Fig.\ \ref{p_wave_B_comp_par}(a)). For $R\approx 1600a_0$ the $s$-wave potential well decreases monotonically for decreasing $R$ which is caused by the level repulsion described in Section \ref{p_wave}. For finite electric field strengths the topology of the $s$-wave PES changes in the sense that the angular positions of minima of the the potential wells ($\text{I}_{\rm{||}}$-$\text{III}_{\rm{||}}$) are shifted to higher $\theta \in [\pi/2, \pi]$ values. This is clearly visible in Fig. \ref{B_60_E_20_exc_10}. This effect is simply explained by the fact the electric field force the electron density to align in its negative $z$-direction which is reflected in a deeper Born-Oppenheimer potential in this region. The radial positions $R_{\rm{I}_{||}}-R_{\rm{III}_{||}}$ of the minima are less strongly affected. With increasing electric field they are transferred to the final values $1750a_0$ ($R_{\rm{I}_{||}}$), $1940a_0$ ($R_{\rm{II}_{||}}$) and $2175a_0$ ($R_{\rm{III}_{||}}$) for $E>80\ \frac{V}{m}$. Furthermore, we see that the larger the radial position $R$ of the considered potential well from the ionic core the larger is the angular distance from $\theta=\pi/2$. This feature can be understood in a semiclassical picture where we compare the Lorenz force $F_{\rm{L}}$ with the electrostatic force $F_{\rm{el}}$ on the electron. Since $F_{\rm{L}}\sim v_{\rm{el}}B\sim \sqrt{1/R-\frac{1}{2n^2}}B$ the Lorentz forces decreases with $R$ while $F_{\rm{el}} \sim E$ remains constant. Therefore the electron density further away from the ionic rubidium core is more strongly affected by the electric field. In general the depth of the potential wells ($\text{I}_{\rm{||}}$-$\text{III}_{\rm{||}}$) strongly varies with the corresponding parameter values. In case of a dominant magnetic field ($B \ge 80$ G, $E \le 60\ \frac{V}{m}$) the wells possess depths up to $100$ MHz. For a dominant electric field ($B \le 40$ G, $E \ge 40\ \frac{V}{m}$) their depths are $200-300$ MHz.   

\begin{figure}[hbt]
\includegraphics[width=0.5\linewidth]{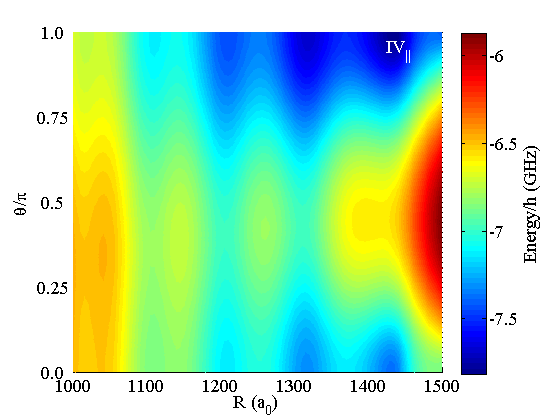}
\caption{(Color online) $p$-wave PES as a function of $(R,\theta)$ ($\theta_{\rm{E}}=\phi_{\rm{E}}=0$, $B=60$ G, $E=20\ \frac{V}{m}$). We observe the global potential minimum at $R=1432a_0,\ \theta=\pi$ providing bound states with a level spacing $10$-$30$ MHz. For a finite electric field strength the reflection symmetry with respect to $\theta=\pi/2$ is broken.}
\label{B_60_E_20_exc_13}    
\end{figure}

In Fig.\ \ref{B_60_E_20_exc_13} we present the $p$-wave PES curve for $B=60$ G, $E=20\ \frac{V}{m}$. It possesses a global equilibrium position at $R=1432a_0, \theta=\pi$ denoted by $\text{IV}_{||}$. As described in Section \ref{p_wave} in case of a pure magnetic field the $p$-wave PES possesses a $\theta \rightarrow \pi-\theta$ symmetry (see Fig. \ref{p_wave_B_comp_par}). A finite electric field along the $z$-axis breaks this symmetry and tends, as above-mentioned, to enhance the electron density in the negative $z-$direction. As a consequence we find in case of the $p$-wave PES the potential minimum at $\theta=\pi$ (Fig.\ \ref{B_60_E_20_exc_13}). The depth of this potential well is approximately $300$ MHz and remains roughly constant for all considered field strengths.
\subsection{Perpendicular field configuration \label{perp_field_conf}}

\begin{figure}[hbt]
\includegraphics[width=0.5\linewidth]{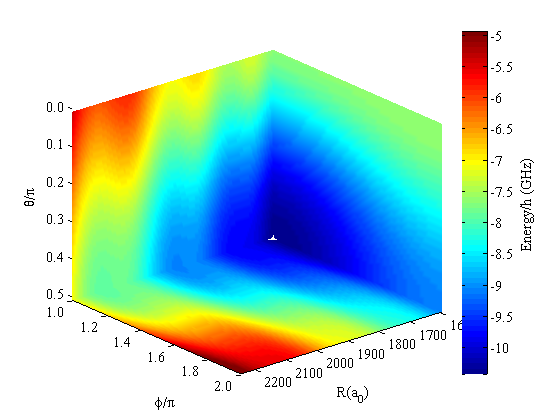}
\caption{(Color online) Three-dimensional $s$-wave PES as a function of $(R,\theta,\phi)$ ($\theta_{\rm{E}}=\pi/2, \phi_{\rm{E}}=0$, $B=100$ G, $E=60\ \frac{V}{m}$). We find two local potential wells at $R=2159a_0$ and $R=1922a_0$, $\theta=\pi/2$ and $\phi=\pi$. For more details an intersection for $\theta=\pi/2,\ \phi=\pi$ is presented in Fig.\ \ref{curves_e_field_var}.} 
\label{B_100_E_60_exc_10_cut}
\end{figure}

\begin{figure}[hbt]  
\includegraphics[width=0.5\linewidth]{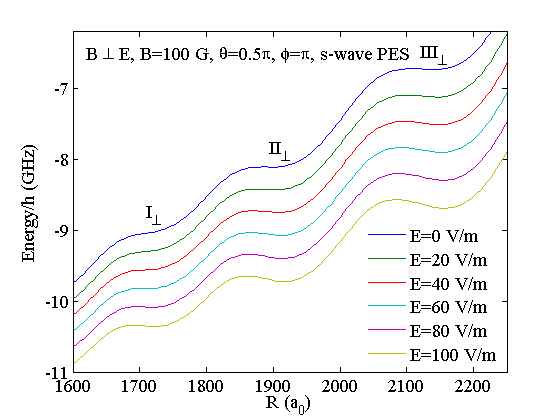}
\caption{(Color online) Intersections through the $s$-wave PES for the perpendicular field field configuration for $\theta=\pi/2,\ \phi=\pi$. The shown cuts are taken for fixed $B=100$ G while the electric field strength $E$ is varied from $E=0$ to $100\ \frac{V}{m}$ in steps of $20\ \frac{V}{m}$. Depending on the electric field strength we find local plateaus / potential wells at $R=1728a_0,\ 1918a_0$ and $2159a_0$ labeled by $\text{I}_{\perp}$, $\text{II}_{\perp}$ and $\text{III}_{\perp}$. With increasing $E$ the plateaus ($B=100$ G) are transferred into local potential wells with depths of $40$ MHz ($\text{I}_{||}$), $100$ MHz ($\text{II}_{||}$) and $140$ MHz ($\text{III}_{||}$).}
\label{curves_e_field_var} 
\end{figure}

Fig.\ \ref{B_100_E_60_exc_10_cut} we shows the three-dimensional $s$-wave PES for the crossed field configuration $B=100$ G, $E=60\ \frac{V}{m}$ as a function of $(R,\theta,\phi)$. Because of the $\theta \rightarrow \pi-\theta$ and $\phi \rightarrow 2\pi-\phi$ symmetries (see Section \ref{method}) we present the potential surface in the range of $0 \le \theta \le \pi/2$ and $0 \le \phi \le \pi$.  We clearly see an oscillating structure with local potential minima aligned into the negative $x$-direction ($\theta=\pi/2,\ \phi=\pi$). This can be understood by the fact the electric field simply deforms the azimuthaly symmetric PES for a finite magnetic field strength in the sense that it forces the electron density to align along the negative $x$-direction. Due to this we obtain molecular states with a well-defined orientation antiparallel to the electric field. In contrast to the parallel field configuration the orientation of these molecular states cannot be tuned by varying the electric and magnetic field strengths. In the considered parameter range tuning the field parameters just changes the depth of the local potential minima at $\theta=\pi/2,\ \phi=\pi$. This feature is shown in detail in Fig.\ \ref{curves_e_field_var} which presents one-dimensional potential cuts for the crossed field configuration. We have fixed the magnetic field strength to $B=B_{\rm{cr}}=100$ G and vary the electric field from $E=0-100$ $\frac{V}{m}$ in steps of $20\ \frac{V}{m}$. As already discussed in Section \ref{p_wave} for $E=0$ we obtain no potential local wells, i.e.\ no bound states are provided for this magnetic field strength. With increasing electric field strength we again obtain local potential wells at the minima positions $R_{\rm{I}_{\perp}}=1728a_0$ $R_{\rm{II}_{\perp}}=1918a_0$ and $R_{\rm{III}_{\perp}}=2159a_0$ respectively. Obviously, these values are very close to those obtained for the parallel field configuration (see previous subsection). Similar to the parallel field configuration we label the wells/plateaus with ($\text{I}_{\rm{\perp}}$), ($\text{II}_{\rm{\perp}}$) and ($\text{III}_{\rm{\perp}}$). The radial equilibrium positions increase with increasing electric field strength up to $1750a_0$ ($R_{\rm{I}_{\perp}}$), $1940a_0$ ($R_{\rm{II}_{\perp}}$) and $2175a_0$ ($R_{\rm{III}_{\perp}}$). As we observe in Fig.\ \ref{curves_e_field_var} the well ($\text{III}_{\rm{\perp}}$) is affected most by the increasing electric field in the sense that its depth increases from $0$ up to $140$ MHz. Similarly the depths of the well ($\text{II}_{\rm{\perp}}$) and ($\text{I}_{\rm{\perp}}$) increase up to $100$ MHz ($\text{II}_{\rm{\perp}}$) and $40$ MHz ($\text{I}_{\rm{\perp}}$) respectively. We therefore conclude that the electric field counterbalances the effect of the $p$-wave interaction and leads to bound states where otherwise none would have existed. This result is reminiscent of an effect already observed for the pure electric field configuration where the electric field stabilizes bound molecular states of the $s$-wave PES \cite{Kurz13} as well. For dominant electric fields in the considered field regime the depths of the potential wells increase up to a value of approximately $300$ MHz.

\begin{figure}[hbt]
\includegraphics[width=0.5\linewidth]{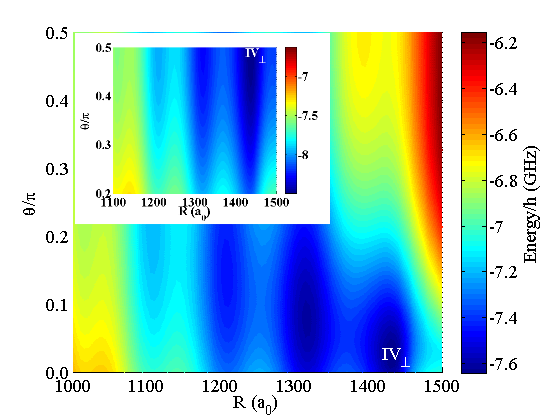}
\caption{(Color online) Two-dimensional $p$-wave PES for $\phi=\pi$ and $0 \le \theta \pi/2$ ($\theta_{\rm{E}}=\pi/2,\ \phi_{\rm{E}}=0$, $B=100$ G, $E=80\ \frac{V}{m}$). The complete PES is a function of $(R,\theta,\phi)$ and possesses a $\theta \rightarrow \pi-\theta$ symmetry. We observe several local potential wells with the energetically lowest labeled by $\text{IV}_{\perp}$. The inset shows the same PES but for $B=0$ G, $E=80\ \frac{V}{m}$. In this case the $p$-wave PES possesses a single potential minimum $\text{IV}_{\perp}$ at $R_{\rm{eq}}=1432a_0,\ \theta=\pi/2,\ \phi=\pi$. }
\label{phi_1_B_100_E_80_exc_13} 
\end{figure}

In Fig.\ \ref{phi_1_B_100_E_80_exc_13} we present a two-dimensional cut defined by $\phi=\pi$ through the $p$-wave PES for the crossed field configuration for $B=100$ G, $E=60\ \frac{V}{m}$. As for the $p$-wave PES in the parallel field configuration, we concentrate on the potential well providing the global equilibrium position. This well is labeled with $\text{IV}_{\perp}$. The radial equilibrium position is again given by $R=1432a_0$ for $\theta=0,\pi$. In case of a pure electric field the single existing  potential minimum is localized at $R=1432a_0,\ \theta=\pi/2, \phi=\pi$ which is shown in the inset of Fig.\ \ref{phi_1_B_100_E_80_exc_13}. By increasing the magnetic field strength the angular equilibrium position is shifted from $\theta_{\rm{eq}}=\pi/2$ to $\theta_{\rm{eq}}=\pi/2 \pm \delta,\ \delta \in(0,\pi/2]$. This means the topology of the $p$-wave PES changes from a single well to a double well PES. For all applied field strengths the depth of the well $\text{IV}_{\perp}$ remains around $300$ MHz.

In Tab.\ \ref{table_2} we summarize the topological properties for both field configurations in the limit of dominant electric and magnetic field strengths.

\begin{table}[t]
\centering
\begin{tabular}{|c|c|c|c|c|c|c|c|c|c|}\hline
 & $\text{I}_{||}$  & $\text{II}_{||}$  & $\text{III}_{||}$ &  $\text{IV}_{||}$ & $\text{I}_{\perp}$  & $\text{II}_{\perp}$  & $\text{III}_{\perp}$ &  $\text{IV}_{\perp}$ \\ \hline
 $B \gg E$ & $1728a_0$,$\frac{\pi}{2}$ & $1918a_0$,$\frac{\pi}{2}$ & $2159a_0$,$\frac{\pi}{2}$ & $1432a_0$,($0,\pi$) & $1728a_0$,$\frac{\pi}{2}$,$\pi$ & $1918a_0$,$\frac{\pi}{2}$,$\pi$ & $2159a_0$,$\frac{\pi}{2}$,$\pi$ & $1432a_0$,$\frac{\pi}{2}$,$\pi$\\ \hline
   $B \ll E$ & $1750a_0$,$\pi$  & $1940a_0$,$\pi$ & $2175a_0$,$\pi$ &$1432a_0$,$\pi$ & $1750a_0$,$\frac{\pi}{2}$,$\pi$ & $1940a_0$,$\frac{\pi}{2}$,$\pi$ & $2175a_0$,$\frac{\pi}{2}$,$\pi$ & $1432a_0$,($0,\pi$),$\pi$ \\ \hline   
\end{tabular}
\caption{Topological properties of the $s$- and $p$-wave PES for both field configurations for dominant magnetic ($B \gg E$) and electric ($B \ll E$) field. The triple $(R_{\rm{eq}},\theta_{\rm{eq}},\phi_{\rm{eq}})$ presents the radial and angular equilibrium positions. In case of azimuthally symmetric PES (parallel fields) only ($R_{\rm{eq}},\theta_{\rm{eq}})$ is provided. In case the considered PES possesses double well character both angular equilibrium positions are presented (e.g.\ ($1432a_0$,($0,\pi$),$\pi$) for $\text{IV}_{\perp}$, $B \ll E$).  }
\label{table_2}
\end{table}
\section{Rovibrational States \label{rovi_states}}
\begin{figure}[hbt]
\includegraphics[width=0.6\linewidth]{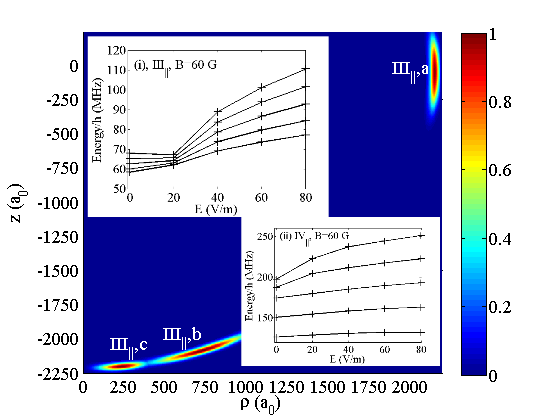}
\caption{(Color online) Scaled probability densities $|F_{00}(\rho,z)|^2$ for rovibrational wave functions. Shown are the ground state probability densities in the potential well ($\text{III}_{\rm{||}}$) for electric field strengths $E=0,20,40\frac{V}{m}$ for the parallel field configuration with $B=60$ G. The densities for the corresponding field strengths are labeled by $a,b,c$. In the inset (i) the rovibrational eigenenergies for the five energetically lowest states in the ($\text{III}_{\rm{||}}$) well are shown with varying electric field strength $E$. The level spacing first decreases and consequently increases up to $10$ MHz. The inset (ii) shows the eigenenergies for the five energetically lowest states in the ($\text{IV}_{||}$) well of the $p$-wave PES. Here the level spacings for the energetically lowest states remain constant and only for the higher excited states the increasing angular confinement causes an increase. We find a level spacing between $10-30$ MHz.}   
\label{density}  
\end{figure}

To analyze the rovibrational states for the parallel field configurations we introduce cylindrical coordinates ($\rho,Z,\phi$) for the parametrization $\epsilon_{||}(\rho,Z)$. We have $[H_{\rm{rv}},L_{\rm{z}}]=0$, which means the azimuthal quantum number $m$ is a good quantum number. With this we write the rovibrational wave function $\phi(\rho,Z,\varphi)=\frac{F_{\nu m}(\rho,Z)}{\sqrt{\rho}}\exp(im\varphi),\ m \in \mathbb{Z},\ \nu \in \mathbb{N}_0$ which transforms the Hamiltonian (\ref{hamrovi}) into 
\begin{eqnarray}
H_{\rm{rv}}&=&-\frac{1}{M}(\partial^{2}_{\rho}+\partial^2_{Z})+\frac{m^2-
1/4}{M\rho^2}+\epsilon_{||}(\rho,Z).\label{ham_rovi_cyl}
\end{eqnarray}
We solve the corresponding Schr\"odinger equation focusing on $m=0$ using a fourth order finite difference method for electric field strengths in the range $0,20,...,80\ \frac{V}{m}$ and $B=60$ G. In Fig.\ \ref{density} we present the ground state probability densities of the local potential well ($\text{III}_{||}$) in cylindrical coordinates ($\rho,Z$). We label the densities according to the applied field strengths $E$ with $a,b,c$. For instance, ($\text{III}_{||}$,c) indicates the probability density of the ground state in the well ($\text{III}_{||}$)  with an applied electric field strength of $E=40\ \frac{V}{m}$. As described in Section \ref{pot_ener_surf} with increasing electric field strength the wells move from the $\theta=\pi/2$ configuration to the $\theta=\pi$ direction. This feature is clearly reflected in the position of the ground state probability densities. For field strengths beyond $E=40\ \frac{V}{m}$ the position of the minimum of the well remains close to $\theta=\pi$, as a result there are no qualitative changes of the corresponding ground state rovibrational probability densities. The latter can be characterized by their radial $\Delta \theta$ and angular extension $\Delta \theta$. typical values observed are of the order of $\Delta R=80a_0$ and $\Delta \theta=250a_0$ ($\text{III}_{||}$,c) up to $650a_0$ ($\text{III}_{||}$,b). We see that with increasing electric field strength from $E=0$ to $20\ \frac{V}{m}$ the angular extension of the rovibrational probability densities in ($\text{III}_{||}$,a) and ($\text{III}_{\rm{||}}$,b) increase as well. For higher field strengths the potential well ($\text{III}_{\rm{||}}$) approaches $\theta=\pi$ and the angular extension of the probability density decreases again. This is clearly visible for ($\text{III}_{||}$,c) in Fig.\ \ref{density} and is caused by the potential term $\frac{-1}{4M\rho^2}$. In the inset (i) of Fig.\ \ref{density} we present the the first five eigenenergies of the rovibrational states for the potential well ($\text{III}_{\rm{||}}$) of the $s$-wave PES, relative to the minimum of the potential well. The level spacing decreases to $2$ MHz for $E=20\ \frac{V}{m}$ which can be explained by the minor decrease of the angular confinement. With increasing $E$ beyond a field strength of $40\ \frac{V}{m}$ the potential gets affected by the centrifugal term in Eq.\ (\ref{ham_rovi_cyl}) and the angular confinement increases. This leads to a larger level spacing up to $10$ MHz as can be seen in the inset (i) of Fig.\ \ref{density}.

For the $p$-wave PES the level spacing is of the order of $10-30$ MHz (see inset (ii) of Fig. \ref{density}). In case of the low-lying states it hardly varies with increasing field strength, only the higher excited states are affected in the sense that their level spacing increases from $5$ to $10$ MHz. This can be explained by the fact that enhancing the electric field strength increases the angular confinement which more strongly affects the higher excited states than the energetically low-lying ones.\\    
In case of the crossed field configuration we have $[H_{\rm{rv}},P_{Y}]=[H_{\rm{rv}},P_{Z}]=0$ where $P_{Y}:\ Y \rightarrow -Y$ and $P_{Z}:\ Z \rightarrow -Z$. Due to these symmetry properties the wave functions $F(\rho,Z,\varphi)$ now obey $F(\rho,-Z,\varphi)=\pm F(\rho,Z,\varphi)$ and $F(\rho,Z,2\pi-\varphi)=\pm F(\rho,Z,\varphi)$. To estimate the rovibrational level spacings we use the fact that the exact potential energy surfaces can be expanded around their equilibrium positions $(R_{\rm{eq}},\theta_{\rm{eq}},\pi)$ as
\begin{eqnarray}
\epsilon_{\perp}(R,\theta,\phi) \approx \epsilon_{\perp}(R_{\rm{eq}},\theta_{\rm{eq}},\pi)+\frac{1}{4}M\omega^{2}_{R}(R-R_{\rm{eq}})^2+\frac{1}{4}M\omega^{2}_{\theta}R^{2}_{\rm{eq}}(\theta-\theta_{\rm{eq}})^2+\frac{1}{4}M\omega^{2}_{\phi}R^{2}_{\rm{eq}}(\phi-\pi)^2 \label{perp_approx}
\end{eqnarray}
\begin{figure}[hbt] 
\includegraphics[width=0.5\linewidth]{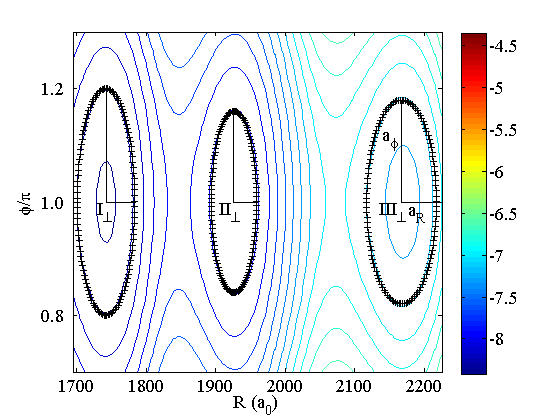}
\caption{(Color online) Contour plot of the $s$-wave PES for $\theta=\pi/2$ with applied field strengths $B=20$ G, $E=60\ \frac{V}{m}$. ($\theta_E=\pi/2$). We clearly see the $\phi=\pi$ reflection symmetry. The ellipses indicated by the black crosses represent approximations to the exact potential surfacea according to Eq.\ (\ref{perp_approx}). From the ellipse parameters we get a level spacing of approximately $130$ MHz for the radial and $5$ MHz for the angular degree of freedom via Eq.\ (\ref{ellipse}).}
\label{contour_perp} 
\end{figure} 
For small extensions in the angular directions $(\theta,\phi)$ the quantities $R_{\rm{eq}}(\phi-\pi)$, $R_{\rm{eq}}(\theta-\theta_{\rm{eq}})$ define together with $R-R_{\rm{eq}}$ a local Cartesian coordinate system. In these coordinates, Eq.\ ($\ref{hamrovi}$) is reduced to three single harmonic oscillators providing level spacings $\omega_{R},\ \omega_{\theta}$ and $\omega_{\phi}$. In Fig.\ \ref{contour_perp} we present such an approximation scheme for the $B=20$ G, $E=60\ \frac{V}{m}$ crossed field configuration for the coordinates ($R,\phi$). We clearly see that the harmonic approximation (crosses) fits the exact potential quite well. From this fit the frequencies of the harmonic oscillator can be extracted
\begin{eqnarray}
\omega_{i}=\sqrt{\frac{4 \Delta V}{M a^{2}_{i}R^2_{\rm{eq}}}},\ \ i=\theta,\phi\ , \ \ \ \  \omega_{R}=\sqrt{\frac{4 \Delta V}{M a^{2}_{R}}}.\label{ellipse}
\end{eqnarray}
where $\Delta V$, $a_i$ and $a_R$ are the energy difference to the potential minimum and the fitted values for the semi-axis of the underlying ellipsoid. For the specific setup we find level spacings of $\omega_{R}=130$ MHz ($\text{I}_{\perp}$), $140$ MHz ($\text{III}_{\perp}$) and $150$ MHz ($\text{II}_{\perp}$). This means each potential well only provides up to one radial excitation. For the angular degrees of freedom we get $\omega_{\phi}\approx \omega_{\theta}=5$ MHz for all three potential wells. In case of the $\text{IV}_{\perp}$ potential well for the $p$-wave PES we obtain level spacings of the order of $\omega_R \approx 200$ MHz in the radial and $\omega_{\theta}=5-20$ MHz, $\omega_{\phi}=5$ MHz in the angular degrees of freedom. 

We remark that for both field configurations and $s$- as well as $p$-wave PES the radial as well as the angular level spacing strongly depends on the applied fields. The general level structure implies a single radial excitation ($130-250$ MHz) with several angular excitations ($5-30$ MHZ) on top. For both field configurations the states in the wells ($\text{I}_{||}$-$\text{III}_{||}$) and ($\text{I}_{\perp}$-$\text{III}_{\perp}$) of the $s$-wave PES possess a finite lifetime due to a tunneling out of the local potential wells. These lifetimes strongly depend on the considered field strengths and we get maximal lifetimes in the order of microseconds.
\section{Molecular Alignment and Orientation \label{mol_align_orien}}
\begin{figure}[hbt]
\includegraphics[width=0.5\linewidth]{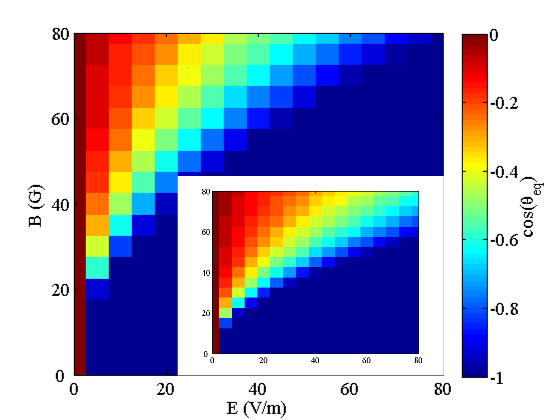}
\caption{(Color online) Orientation $\langle \cos(\theta) \rangle_{\phi} \approx \cos(\theta_{\rm{eq}})$ of the ground state in the potential well ($\text{II}_{||}$) (main figure) and ($\text{I}_{||}$) (inset) belonging to the $s$-wave PES. For pure electric/magnetic fields the internuclear axis is oriented in an antiparallel/perpendicular configuration. By varying the field strengths the orientation can be tuned. For fixed magnetic field strength the antiparallel configuration for the well ($\text{II}_{||}$) is achieved for lower electric field strengths than for ($\text{I}_{||}$).} 
\label{orientation}
\end{figure}

In Section \ref{pot_ener_surf} we presented the possibility to vary the topology of the molecular PES via tuning of the electric and magnetic fields. Obviously, this provides the possibility to control the molecular orientation and alignment. To quantify the orientation and alignment in case of the parallel field configuration we have to analyze the expectation value $\langle \cos(\theta) \rangle_{\phi}$ and the variance $\Delta \cos(\theta) =\sqrt{\langle \cos(\theta)^2 \rangle_{\phi}-\langle \cos(\theta) \rangle^{2}_{\phi}}$. The expectation values $\langle ... \rangle_{\phi}$ are taken with respect to the rovibrational state $\phi(\kd{R})$ for the ground states in the potential wells ($\text{I}_{||}$), ($\text{II}_{||}$) and ($\text{III}_{||}$). The closer the absolute value of $\langle \cos(\theta) \rangle_{\phi}$ is to one, the stronger is the orientation of the state into the $z$-direction and the closer $\Delta \cos(\vartheta)$ is to zero, the stronger is the alignment of the state. We consider the ground state probability densities in the single wells to be strongly localized such that we can approximate the expectation values according to $\langle \cos(\theta) \rangle_{\phi} \approx \cos(\theta_{\rm{eq}})$ and $\langle \cos(\theta)^2 \rangle_{\phi} \approx \cos(\theta_{\rm{eq}})^2$ where $\theta_{\rm{eq}}$ denotes the angular equilibrium position of the underlying potential well (see Fig.\ \ref{density}). In this approximation we get for the variance $\Delta \cos(\theta) \approx 0$, which means that the degree of alignment is perfect. In Fig.\ \ref{orientation} we present the dependence of $\cos(\theta_{\rm{eq}})$ of the ground state state of the potential well ($\text{II}_{||}$) on the applied field. The inset in this figure shows the same analysis but for the ($\text{I}_{||}$) well. We see that for pure magnetic and pure electric fields the state is oriented in a perpendicular (red region) and antiparallel (blue region) configuration, respectively. For both potential wells we find a crossover regime (yellow region) between these two configurations. We see that for fixed magnetic field strength the antiparallel configuration for the well ($\text{II}_{||}$) is achieved for lower electric field strengths than for ($\text{I}_{||}$). This can be explained by the fact that the electric field stronger affects the states in the well ($\text{II}_{||}$) as compared to ($\text{I}_{||}$) (see Section \ref{pot_ener_surf}).
\begin{figure}[hbt]
\includegraphics[width=0.5\linewidth]{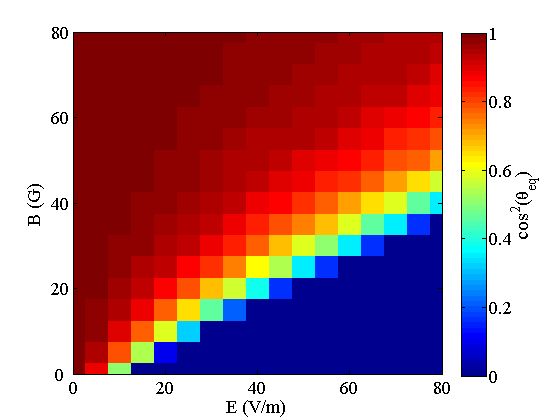}
\caption{(Color online) Alignment $\langle \cos(\theta)^2 \rangle_{\phi} \approx \cos(\theta_{\rm{eq}})^2$ of the ground state in the potential well $\text{IV}_{\perp}$ belonging to the $p$-wave PES. For pure electric/magnetic fields the molecule is anti-/aligned with respect to the magnetic field.}
\label{alignment}
\end{figure}

In case of the crossed field configuration the rovibrational Hamiltonian (\ref{ham_rovi_cyl}) possesses a $P_z$ reflection symmetry i.e. we get \ $\langle \cos(\theta) \rangle_{\phi}=0$. In this case the molecular alignment is quantified by $\langle \cos^2(\theta) \rangle \approx \cos^2(\theta_{\rm{eq}}),\ \theta_{\rm{eq}} \in [0,\pi/2]$ for ground rovibrational states. For the $s$-wave PES we have molecular states with a well-defined perpendicular configuration of the internuclear axis with respect to the $z$-axis. Here we have $\cos^2(\theta_{\rm{eq}})=0$, which means they are antialigned with respect to the $z$-axis. For the $p$-wave PES and finite magnetic field strength we have potential surfaces with double well character. In this case the rovibrational states are delocalized over the double wells. In Fig.\ \ref{alignment} we present the field-dependent alignment of the ground states in the corresponding $\text{IV}_{\perp}$ potential well. We see that for pure electric and magnetic field configuration we have perfectly (anti)aligned molecular states. Similar to the molecular orientation for the parallel field configuration we find a crossover regime (yellow region) where the alignment changes from antialigned (blue region) to aligned states (red region). 
\section{Electric dipole moment \label{elec_mag}}
Due to the impact on the molecular configuration the electric dipole moment can be readily tuned by changing the field strengths and specific field configuration. Let us analyze the dipole moments along the internuclear axis in the following
\begin{eqnarray}
D_{\rm{el}}&=&\langle \psi(\kd{r};\kd{R}_{\rm{eq}};\kd{B},\kd{E})|\kd{n} \cdot \kd{r}| \psi(\kd{r};\kd{R}_{\rm{eq}};\kd{B},\kd{E}) \rangle\label{elec_dipol_formula},
\end{eqnarray}
where $\kd{n}$ denotes the unit vector along the internuclear axis.

\begin{figure}[hbt]
\begin{minipage}{0.48\textwidth}
\includegraphics[width=\linewidth]{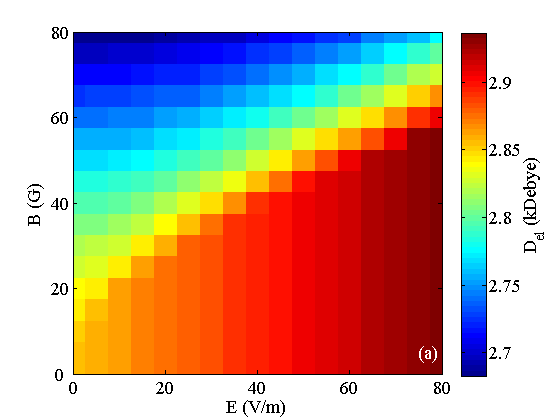}
\end{minipage}
\begin{minipage}{0.48\textwidth}
\includegraphics[width=\linewidth]{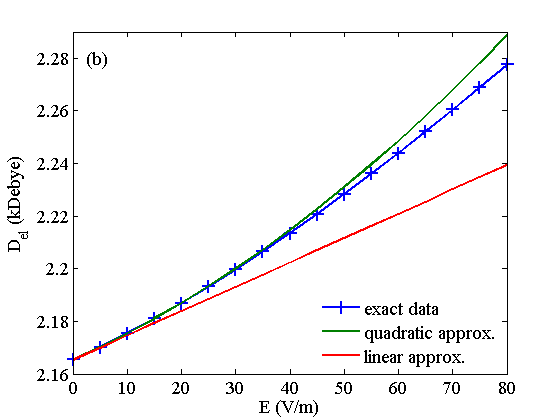}
\end{minipage}
\caption{(Color online) (a) shows the electric dipole moment $D_{\rm{el}}$ in the direction of the internuclear axis for the $s$-wave PES for the parallel field configuration. Using the mean value $\bar{R}_{\rm{eq}}=1739a_0$ the semiclassical approximation $D_{\rm{el}}\approx \bar{R}_{\rm{eq}}-\frac{n^2}{2}$ gives $D_{\rm{el}}\approx 2.85$ kDebye. (b) shows the electric dipole moment for the $p$-wave PES. In addition we show a comparison with the semi-analytic expansion (\ref{state_exp}). The red respectively green curve indicate a linear (O$(E)$) and quadratic approximation (O($E^2$)).}
\label{dipol}
\end{figure}

In Fig.\ \ref{dipol} we show the electric dipole moment for the $s$-wave PES for the parallel field configuration as a function of $E$ and $B$. We observe that with increasing $B$ the dipole moment decreases while it increases for increasing $E$. This can be understood by the fact that in the absence of any contact interaction and $B \not=0,\ E=0$ the reflection operations $P_x \bigotimes P_y$ and $P_z$ are exact symmetries of the Hamiltonian (\ref{hamelec}). In the presence of the neutral perturber and $B=0,\ E=0$ the mixing of degenerate Rydberg states leads to an electric dipole given by the semiclassical approximation $D_{\rm{el}} \approx R_{\rm{eq}}-\frac{n^2}{2}$ for a purely $s$-wave interaction dominated PES. However, with increasing magnetic field the magnetic field terms become dominant and the corresponding symmetry properties get imprinted in the quantum states \cite{Lesanovsky07}. For a pure strong magnetic field case the $s$-wave PES is approximately dominated by the $| 35,34,-34 \rangle$ hydrogen state which explains the decrease of the electric dipole moment. In case of an increasing electric field the electron cloud is more aligned into the negative field direction which causes the increase of $D_{\rm{el}}$.

Next we perform some (semi)analytical analysis to estimate the electric dipole moment. First we check the validity of the semiclassical approximation for finite electric and magnetic field strengths. As discussed in Section \ref{pot_ener_surf} for the considered field regimes the radial positions of the potential wells is only to a minor extent affected by the external fields (see Tab.\ \ref{table_2}). Therefore, we estimate the electric dipole moment as $\bar{R}_{\rm{eq}}-\frac{n^2}{2}$ where $\bar{R}_{\rm{eq}}$ denotes the mean value of the minimal and maximal radial positions for a considered potential well for varying field strength. For instance, $R_{\rm{eq},I_{||}}=1728a_0$ for $B =80$ G, $E=0$ and $R_{\rm{eq},I_{||}}=1750a_0$ for $B=0$ G, $E=100\ \frac{V}{m}$  which gives  $\bar{R}_{\rm{eq},I_{||}}=(1728a_0+1750a_0)/2=1739a_0$ and an approximate dipole moment of $D_{\rm{el}} \approx 2.85$ kDebye. This corresponds to a relative deviation of $5 \%$ compared to the exact result, which means that in the considered parameter regime this simple estimate is quite accurate.

As discussed in Section \ref{pot_ener_surf} in case of the $p$-wave PES its equilibrium position remains constant ($R_{\rm{eq}}=1432a_0,\ \theta_{\rm{eq}}=\pi$) with respect to a variation of the field strengths. In particular, in Section \ref{p_wave} we have shown that the corresponding electronic eigenvector $| \psi(\kd{r};\kd{R}_{\rm{eq}};B,E) \rangle$ is independent of the applied magnetic field. Due to this we can reduce the analysis of the dipole moment of the $p$-wave state to an arbitrary value of $B$ which we choose to be $B=0$ G. In Fig.\ \ref{dipol}(b) we show the electric dipole moment for the $p$-wave PES for $B=0$ G as a function of $E$. With increasing electric field strength $D_{\rm{el}}$ grows quadratically. To verify this we present a corresponding semi-analytical result for the electric dipole moment where we expanded the state $| \psi(\kd{r};\kd{R}_{\rm{eq}};B=0,E) \rangle$ in a perturbative series up to the $O(E^2)$:
\begin{eqnarray}
|\psi(\kd{r};\kd{R}_{\rm{eq}};0,E)\rangle=|\psi_0(\kd{r};\kd{R}_{\rm{eq}})\rangle+E\sum_{n \not =0}C^{(1)}_{n}|\psi_n(\kd{r};\kd{R}_{\rm{eq}})\rangle +E^2\sum_{n \not =0}C^{(2)}_{n}|\psi_n(\kd{r};\kd{R}_{\rm{eq}})\rangle .\label{state_exp}
\end{eqnarray} 
In this expansion $|\psi_n(\kd{r};\kd{R}_{\rm{eq}})\rangle$ indicate the field-free electronic states and $C^{(1,2)}_{n}$ are the expansion coefficients given by standard perturbation theory \cite{sakurai}. Inserting this ansatz into (\ref{elec_dipol_formula}) and keeping terms up to $O(E)$ and $O(E^2)$ we obtain the linear term (red) and quadratic term (green) approximations according to Fig.\ \ref{dipol}(b). We see that the exact data (blue line and crosses) are well approximated by the quadratic approximation. The semiclassical approximation gives a result of $D_{\rm{el}}=2.08$ kDebye which deviates from the obtained data by $10 \%$. The larger deviation compared to the $s$-wave state can be explained by the fact that the semiclassical approximation is originally derived for $s$-wave interactions in the absence of any fields. Although we expect the $p$-wave state to possess a strong $s$-wave character far away from the region of avoided crossings ($R < 1450a_0$) for the region of the localized potential well ($R_{\rm{eq}}=1432a_0$) the $p$-wave character still provides a substantial contribution which explains the less accurate result for the resulting electric dipole moment.\\   
For the crossed field configuration the electric dipole moments of the $s$-wave PES potential wells ($\text{I}_{\perp}$-$\text{III}_{\perp}$) show a qualitatively similar behavior as their counterparts ($\text{I}_{||}$-$\text{III}_{||}$) in case of the parallel configuration. For all potential wells we find a decreasing dipole moment for increasing magnetic field strength and an increase of $D_{\rm{el}}$ for an increasing electric field strength. In Tab.\ \ref{table} we present the minimal ($D_{\rm{el,min}}$) and maximal ($D_{\rm{el,max}}$) value of the electric dipole moment (\ref{elec_dipol_formula}) for the wells ($\text{I}_{\perp}$-$\text{III}_{\perp}$). As for the parallel field configuration we compare these results with the semiclassical approximation $\bar{R}_{\rm{eq}}-\frac{n^2}{2}$. With a maximal deviation of $1$-$5 \%$ the exact results are reproduced satisfactorily.    
\begin{table}
\centering
\begin{tabular}{|c|c|c|c|}\hline
    & $D_{\rm{el,min}}$  & $D_{\rm{el,max}}$  & $\bar{R}_{\rm{eq}}-\frac{n^2}{2}$  \\ \hline
   $\text{I}_{\perp}$ & 2.69 & 2.94 & 2.86 \\ \hline
   $\text{II}_{\perp}$ & 3.25  & 3.39 & 3.34\\ \hline
   $\text{III}_{\perp}$ & 3.92 & 3.98 & 3.95 \\ \hline
\end{tabular}
\caption{Minimal ($D_{\rm{el,min}}$) and maximal ($D_{\rm{el,max}}$) electric dipole moment for the potential wells ($\text{I}_{\perp}$-$\text{III}_{\perp}$). The minimal (maximal) values are taken for $B=80$ G, $E=0\ \frac{V}{m}$ ($B=0$ G, $E=80\ \frac{V}{m}$). For comparison we present the semiclassical approximation $\bar{R}_{\rm{eq}}-\frac{n^2}{2}$.}
\label{table}
\end{table}
\section{Conclusions \label{conclusion}}
The recent experimental progress in preparing, detecting and probing the properties of non-polar ultralong-range Rydberg molecules \cite{Krupp14, Bendkowsky09,Pfaunat11,Bendkowsky10,Balewski13} has opened the doorway for a variety of possibilities to create novel species where atoms, molecules and even mesoscopic quantum systems are bound to Rydberg atoms. Therefore, the understanding and control of the properties of these hybrid Rydberg systems is of essential importance. Due to the high sensitivity of the weakly bound Rydberg electron the primarily choice to obtain electronic as well as rovibrational control is the application of external fields.

In this work we have therefore explored the effect of combined electric and magnetic fields on the polar high angular momentum molecular states for a parallel as well as a crossed field configuration. Taking into account both $s$- and $p$-wave interactions it turns out that for a pure magnetic field configuration strong level repulsion causes the potential wells which provide the trilobite states in the field-free case \cite{Greene00} to vanish. For this PES beyond a critical field strength of around $100$ G no bound states are provided anymore. For finite field strengths the angular degrees of freedom are converted from rotational to vibrational degrees of freedom, thereby rendering the field-free potential energy curve into a two- and three-dimensional energy surface for parallel and crossed field configurations, respectively. We obtain oscillatory potential curves with localization in the radial and angular degrees of freedom with depths up to hundreds of MHz providing a rich topology depending on the specific degree of electronic excitation and field configuration. The resulting rovibrational level spacings are in the order of several MHz. The parallel as well as the crossed field configuration provide unique ways to control the topology of the adiabatic potential energy surfaces. This directly leads to the possibility to control molecular orientation and alignment for the parallel and crossed field configuration, respectively. For instance, for parallel fields the molecular orientation can be tuned from a perpendicular to an antiparallel configuration by varying applied field strengths. In case of crossed fields the molecular alignment can be changed between an aligned and anti-aligned configuration with respect to the magnetic field. In addition, the topological control of the PES provides the possibility of directly controlling the electric dipole moment as well. Apart from numerical results we have provided also a semiclassical estimate and perturbative analysis of the electric dipole moment.

The plethora of interesting effects of high-$\ell$ ultralong-range Rydberg molecules in external fields keeps this particular species a promising candidate for future investigations. Because of its high sensitivity to small field strengths it is worth studying the dependence of molecular properties like electric and magnetic polarizabilities and susceptibilities. In case of Rydberg atoms these quantities strongly depend on the Rydberg excitation. 

\section{Acknowledgment}
We thank the Initial Training Network COHERENCE of the European Union FP7 framework for financial support. 
\end{document}